\documentclass[debug]{rmaa}

\usepackage{txfonts}

\usepackage{paralist}

\usepackage{psfrag,color}

\usepackage[latin1]{inputenc}

\title{Interpretation of optical and IR light curves for transitional 
disks candidates in NGC 2264 using the extincted stellar radiation and the 
emission of optically thin dust inside the hole }
 
\author{
   E. Nagel,\altaffilmark{1} 
   F. Guti\'errez-Canales,\altaffilmark{1}
   S. Morales-Guti\'errez,\altaffilmark{1}
   and A.P. Sousa\altaffilmark{2}}

\altaffiltext{1}{Departamento de Astronom\'\i{}a, Universidad de
Guanajuato, M\'exico.}

\altaffiltext{2}{Institut de Plan\'etologie et d'Astrophysique de Grenoble, 
Universit\'e Grenoble Alpes, France.}

\shortauthor{Nagel, et al.}
\shorttitle{Interpretation of transitional disk candidates light curves}

\fulladdresses{
\item Fernando Guti\'errez-Canales, Sebasti\'an Morales-Guti\'errez and
  Erick Nagel: Departamento de Astronom\'\i{}a, 
  Universidad de Guanajuato, 
  Callej\'on de Jalisco S/N, 36240 Guanajuato, Guanajuato, M\'exico
  (e.nagel@ugto.mx).
\item Alana P. Sousa: Universit\'e Grenoble Alpes, IPAG, F-38000 Grenoble, France.}

\listofauthors{E. Nagel, F. Guti\'errez-Canales, S. Morales-Guti\'errez \& A.P. Sousa}
\indexauthor{Nagel, E..}
\indexauthor{Guti\'errez-Canales, F.}
\indexauthor{Morales-Guti\'errez, S.}
\indexauthor{Sousa, A.P.}

\abstract{In the stellar forming region NGC 2264 there are objects catalogued as hosting a transitional disk according to spectra modeling. Four members of this set have optical and infrared light curves coming from the \textit{CoRoT} and \textit{Spitzer} telescopes. In this work, we try to simultaneously explain the light curves using the extinction of the stellar radiation and emission of the dust inside the hole of a transitional disk. For the object Mon-296, we were successful to do this. However, for Mon-314, and Mon-433 our evidence suggests that they host a pre-transitional disk. For Mon-1308 a new spectra fitting using the 3D radiative transfer code Hyperion, allow us to conclude that this object host a full-disk instead of a transitional disk. This is in accord to previous work on Mon-1308 and with the fact that we cannot find a fit of the light curves only using the contribution of the dust inside the hole of a transitional disk.}

\resumen{En la regi\'on de formaci\'on estelar NGC 2264 hay objetos catalogados como anfitriones de discos transicionales de acuerdo al modelaje de su espectro. Cuatro miembros de este conjunto tienen curvas de luz en el \'optico y en el infrarrojo provenientes de los telescopios \textit{CoRoT} y \textit{Spitzer}. En este trabajo, tratamos de simult\'aneamente explicar las curvas de luz usando extinci\'on de la radiaci\'on estelar y emisi\'on del polvo dentro del agujero de un disco transicional. Para el objeto Mon-296, fuimos exitosos. Sin embargo, para Mon-314 y Mon-433 nuestra evidencia sugiere que contienen un disco pre-transicional. Para Mon-1308 un nuevo ajuste del espectro usando el c\'odigo 3D de transferencia radiativa Hyperion, nos permite concluir que este objeto tiene un disco completo en lugar de un disco transicional. Esto coincide con trabajo previo sobre Mon-1308 y 
tambi\'en con que no podemos encontrar un ajuste de las curvas de luz solo usando la contribuci\'on del polvo dentro del agujero de un disco transicional.  }

\addkeyword{dust, extinction}
\addkeyword{protoplanetary disks}
\addkeyword{Stars: pre-main sequence}

\begin{document}
\maketitle

\section{Introduction}
\label{sec:intro}

Spectral and photometric variability of young stellar objects (YSOs) is the usual outcome of multiwavelength campaigns \citep{2014AJ...147..83,2015AJ...149..130,2016AJ...151..60,2014AJ...147..82,2011ApJ...733..50}. For young stars in NGC 2264, 
\citet{2014AJ...147..83} extract the accretion burst dominated light curves (lcs), \citet{2015AJ...149..130} show the short-duration periodic flux dips in the lcs and \citet{2016AJ...151..60} present the stochastically varying lcs.
\citet{2014AJ...147..82} extract optical and infrared (IR) lcs from the \textit{Spitzer} and \textit{CoRoT} telescopes for 162 classical T Tauri stars (CTTSs) where flux variations are clearly detected. They catalogued them into seven distinct classes describing multiple origins of young star variability: circumstellar obscuration events, hot spots, accretion bursts and structural changes in the inner disk. Focusing at $3.6$ and $4.5\mu$m, lcs of hundreds of objects in the Orion Nebula Cluster, \citet{2011ApJ...733..50}
found variability that can be interpreted by processes occurring in the disk like density structures intermittently blocking our line of sight. From this and others studies is extracted the label "dippers" for the objects showing changes in the flux that can be explained by circumstellar material crossing the line of sight directed towards the object.
\citet{1999A&A...349..619} refer to the prototypical dipper AA Tau interpreting the lcs by asymmetries at the inner edge of the dusty disk where a magnetically induced warp is formed. This object shows a sudden dimming in 2011 that can be interpreted with the extinction produced by an overdense region orbiting around the star 
\citep{2013A&A...557..A77}. 
\citet{2010A&A...519..A88} use observations of the \textit{CoRoT} telescope to search for AA Tau type like objects in NGC 2264. They conclude that the dipper objects are common because the frequency is $\sim 30$ to $40\%$ in YSOs with dusty disks.

Identification of dippers \citep{2017ApJ...848..97} and its interpretation locating material in the innermost regions of the disk 
\citep{2017MNRAS.470..202,2020A&A...643..A157} is a key issue to characterize
the interaction of the magnetosphere and the disk.
We need a reasonable amount of material in the accretion streams to account for the dipper behavior \citep{2017MNRAS.470..202}. However, the weak accretion signatures of the transitional disks (TDs) in the sample of dippers in 
\citet{2016ApJ...816..69} is enough to interpret the 
variability of the lcs with the extinction of the material in the innermost part of the disk which is interacting with the magnetospheric lines. 

\citet{2016ApJ...816..69} interpret the lcs of the ten objects in their sample using three different mechanisms: occulting inner disk warps, vortices caused by the Rossby Wave Instability (RWI) and transiting circumstellar clumps. Warps require the presence of material in the innermost region of the circumstellar environment which is revealed by strong accretion signatures. The RWI is responsible for forming non-axisymmetric structures \citep{2014FIDyR.46..041401} as vortices 
\citep{2010A&A.516..A31} which explains shallow, short-duration and periodic dippers in the \citet{2016ApJ...816..69}'sample. Transiting circumstellar clumps can explain the lc of the evolved disk in EPIC 205519771 because the lc is aperiodic and the accretion signatures are weak. The few days timescale for the variations in any of these objects leads to assume that any mechanism requires material in the innermost circumstellar zones. 
The interpretation of lcs with low-periodicity indicates that the explanation should include the effects of the highly dynamic environment close to the star. For different campaigns in a subsample of dippers in \citet{2015A&A...577..A11}, there is a change between unstable and stable accretion regimes \citep{2016MNRAS.459..2354}, affecting the mass accretion rate towards the star, $\dot{M}$ \citep{2008MNRAS.386..673}, in this way shaping the behavior of the lcs.

The concept of pre-transitional (PTDs) and TDs have a recent presence in the discussion of YSOs. Using sub-millimeter observations, \citet{2011ApJ...732..42} observed $12$ TDs with cavities in the range from $15$ to $73$AU. 
\citet{2010ApJ...717..441,2011ApJ...728..49} favor its existence modeling the SEDs including the presence of an inner disk component and emission of dust coming from the gap or hole. From this modeling, \citet{2010ApJ...717..441} catalogued LkCa 15, UX TauA, ROX44 as hosting PTDs but GM Aur and DM Tau host TDs. The modeling requires some optically thin dust in the hole of the disk associated for GM Aur but for DM Tau the hole is empty of grains, as previously interpreted by \citet{2005ApJ...630..L185}. The variability of PTDs is interpreted in the sample of 
\citet{2011ApJ...728..49} by changes of $\sim 20\%$ in the inner disk wall height which they associate to a warp. For the TD in GM Aur, the variability between two campaigns is explained with changes in the inner edge height of the disk at $23$AU from $2.9$ to $3.2$AU. The absence of variability for DM Tau is interpreted using the absence of dust within the hole of its TD as an argument to justify the non-existence of a mechanism to explain the variability occurring at this timescale. For GM Aur, the model by \citet{2015ApJ...805..149} requires changes in $\dot{M}$ associated to inhomogeneities in the inner disk as the process explaining ultraviolet, optical and near-infrared (near-IR) observations. 
\citet{2017RMxAA...53..227} also model this object but using the intermittent formation of a sublimation wall associated to accumulation of matter as the physical mechanism to explain variability in the SpeX spectrum. Both analysis point out the multiplicity of ways to explain this kind of objects but restrict the structures formed in the inner region as the relevant aspect to focus on.

In YSOs, one way to interpret optical and infrared lcs is by means of the dust in the disks surrounding them. In many cases, along with the lcs, their Spectral Energy Distributions (SEDs) are the only sources of information coming from them, thus when images are not available, the physical characteristics of these systems should be 
only interpreted by modeling their fluxes or using selection criteria defined by different ranges in some phometric colors and spectral indices \citep{2009A&A...504..461,2010ApJ...718..1200,2010ApJ...712..925,2010ApJ...708..1107}. 

What helps to choose an adequate structure around each object comes from the distinction between a full and a TD shown as a 
different signature in the SED but also its effect on the optical lc.
During the last years using the available new facilities, images of TDs are obtained for a few systems. These images point out that the structure
is complex, presenting vortices and spirals as in the TD HD 135344B 
\citep{2012ApJ...748..L22,2016ApJ...832..id178} with a bias towards large cavities 
\citep{2018ApJ...854..id177}. The formation of spirals is justified with spiral density wave theory in \citet{2012ApJ...748..L22}. Note that the stellar radiation produces a puffed up inner rim that occults some disk regions from this radiation, clearly changing the radial temperature profile and affecting the formation of structures \citep{2010ARA&A...48..205}. 
\citet{2012ApJ...748..71} explains observed variability in the IR range with changes in the inner disk structure, i.e. scale height fluctuations.
We summarize this saying that a physically correct model should be able 
to explain many sources of information: photometry, spectra and images. These ideas highlights the importance of being able to identify which kind of disk we are dealing with. 
However, due to the difficulty to achieve enough spatial resolution to obtain images for many of the objects even in close stellar forming regions, the interpretation leads to degeneracy, thus at most we can find models consistent with the observations. 

In this work, we are focusing on TDs; the interpretation of the information coming from the \textit{Spitzer Space Telescope} and other facilities allow 
to the astronomy community to be confident that TDs are out there
\citep{2007ApJ...670..L135,2008ApJ...689..L145,2014PPVI...497-520}. Surveys of
objects allow a characterization, for instance the observed $\dot{M}$ estimated using the UV excesses caused by the magnetospheric streams falling towards the stellar surface can be explained when these estimates are compared with the value for $\dot{M}$ in classical T Tauri stars \citep{2007MNRAS.378..369,
2015MNRAS.450..3559}. The disk mass ($M_\mathrm{d}$) is correlated with 
$\dot{M}$ either in full disks \citep{2016A&A...591..L3} or in TDs
\citep{2012MNRAS.426..L96,2015MNRAS.450..3559}. This lead us to conclude that 
an estimated $\dot{M}$ using detected UV excesses in TDs is a key piece of
information to guarantee the presence of gas in the hole \citep{2014A&A...568..A18}.
The dust attached to it is responsible to shape lcs by extinction of the stellar radiation. The analysis of lcs by \citet{2016ApJ...816..69} point out occulting inner disk warps and transiting circumstellar clumps as possible processes explaining the observations. For the TD candidates analyzed in our work, a non-zero $\dot{M}$ guarantee that there is gas in the inner region of the disk. Assuming that the dust is attached to the gas,
the previous mechanism or any other that uses dust in the innermost region of the disk either as optically thin dust inside the cavity in a TD or an optically thick dusty ring in the PTD case are a plausible scenario to explain the lcs. 
   
For the modeling is usually assumed that the star is not variable in the timescale of the physical processes included to model the lc variability such that the asymmetry of the dust distribution leads to the features 
observed. The main contributor to the optical emission is the star, this means
that the deepness of the signal in the optical lc is given by the amount of 
dust eclipsing the star. A reservoir of optically thin dust prone to extinct
the star is found in the inner hole of TDs. 

Specifically, we are focusing on the sample of TD candidates in the NGC 2264 stellar forming region presented in \citet{2019A&A...629..A67}. 
From this sample we choose the objects that have contemporaneous optical and 
infrared (IR) lcs from \textit{CoRoT} and \textit{Spitzer} respectively \citep{2015A&A...577..A11}, 
in order to check the effect on the lcs caused by the material in the disk 
hole. These objects are Mon-296, Mon-314, Mon-433 and Mon-1308. We focus the study on the YSO Mon-1308 because for this system, we have
two different scenarios proposed. The first one is analysed in
\citet{2019A&A...625..A45} where they simultaneously explain \textit{CoRoT} and 
\textit{Spitzer} lcs using a full disk as the optically thick structure responsible to 
shape the optical lc because it occult sections of the stellar surface and also
it is responsible to shape the IR lc using the disk emission. The second one is
presented in \citet{2019A&A...629..A67} where they modeled the SED of a 
sample of objects using 3 possible cases: a full disk, no disk and a 
TD. For Mon-1308, their best fit is a TD 
with a $20.18$AU hole which is completely different to the full disk structure required
in \citet{2019A&A...625..A45}. The aim of this work is to complement the analysis in
\citet{2019A&A...625..A45} that search an explanation of the optical and IR
lcs of Mon-1308 using a full disk but extend the analysis using a TD instead. 

As a complementary analysis, we repeat the steps applied to Mon-1308 for
the other 3 TD candidates that also share contemporaneous
IR and optical lcs. From the whole set, we point out differences in the tuning of
the modeling required to look for the interpretation of the lcs of systems 
that have a range of hole sizes spanning from $0.12$AU to $20.18$AU. In the analysis, we should not forget that not all the lcs are periodic and not always there is a high resemblance between the IR and the optical lcs.

In \S~\ref{sec:sample} we present the details of the sample studied, in
\S~\ref{sec:modeling} is explained the model we are working with, in
\S~\ref{sec:results} the modeled lc for the 4 objects studied is presented
and finally in \S~\ref{sec:discussion} and \S~\ref{sec:conclusions} the discussion and conclusions are shown.

\section{Sample of objects studied}
\label{sec:sample}

The sample of objects studied are the ones that belong to the 
sample of \citet{2015A&A...577..A11} for objects having AA-Tau like 
and variable extinction dominated lcs and to the sample of TD candidates 
in \citet{2019A&A...629..A67}. The hypothesis for 
this set of objects is that the dust inside the hole is relevant to 
explain both the lcs in the optical and in the IR. The objects
are: Mon-296, Mon-314, Mon-433, and Mon-1308. The hole size $R_\mathrm{H}$,
the stellar mass $M_{\star}$, the stellar radius $R_{\star}$, the 
stellar temperature $T_{\star}$, the minimum ($\dot{M}_\mathrm{min}$),
the maximum ($\dot{M}_\mathrm{max}$) and the observed disk mass accretion
rate ($\dot{M}_\mathrm{obs}$) are presented in Table~\ref{tab:observed_param}.
Otherwise mentioned, these parameters come from \citet{2014A&A...570..A82}.
The mass accretion rate comes from two estimates: (u-g) and (u-r)
color excesses modeling. $\dot{M}_\mathrm{obs}$ is taken from the first estimate.
$\dot{M}_\mathrm{min}$ and $\dot{M}_\mathrm{max}$ are the minimum and maximum values
coming from both estimates including the errors.
Notice that the range of $R_{H}$ span from $0.12$AU in Mon-296 to $20.18$AU 
in Mon-1308, a difference of more than two orders of magnitude; allowing to 
analyse the effect of this parameter in the lcs.

\begin{table}[!t]\centering
  \setlength{\tabnotewidth}{0.5\columnwidth}
  \tablecols{8}
  \setlength{\tabcolsep}{2.0\tabcolsep}
  \caption{Observed parameters}
 \begin{tabular}{cccccccc}
    \toprule
    Object & $R_{\star}(R_{\odot})$ & $M_{\star}(M_{\odot})$ & 
    $T_{\star}(\,\mathrm{K})$ & $\dot{M}_\mathrm{min}^a$ &
    $\dot{M}_\mathrm{obs}^a$ & $\dot{M}_\mathrm{max}^a$ & $R_\mathrm{H}(AU)$ \\
    \midrule
     Mon-296  & $1.71$ & $1.42$ & $4950$ &  & $1.73$ &  & $0.12$  \\
     Mon-314  & $1.35$ & $0.29$ & $3360$ & $5.75$ & $5.75$ &  $13.5$ & $3.58$  \\
     Mon-433  & $0.99$ & $0.44$ & $3680$ & $1.38$ & $2.04$ &  $2.51$ & $7.39$  \\
     Mon-1308 & $1.59$ & $0.63$ & $3920$ & $5.62$ & $8.51$  &  $28.8$ & $20.18$ \\
    \bottomrule
    \tabnotetext{a}{All the $\dot{M}$ are given in units of $10^{-9}\,\mathrm{M_{\odot}yr^{-1}}$.}
  \end{tabular}
  \label{tab:observed_param}
\end{table}

The analysis of the periodicity of the lcs for the objects in the
sample was done by \citet{2014AJ...147..82} using as a starting point the 
auto correlation function (ACF) defined and used to calculate the
rotational period for a sample of M dwarfs by \citet{2013MNRAS.432..1203}.
Applied to a time series, the maximum of the ACF gives the lapse of 
time elapsed between two points for which the signal is the most
correlated. In order to confirm that the selected period really 
corresponds to the main period of the data, \citet{2014AJ...147..82} 
compute a Fourier transform periodogram and search for peaks within 
$15\%$ of the frequency associated to the period obtained. From this 
analysis, the objects in our sample show the periods given in 
Table~\ref{tab:periods}.

Two of the systems show a clear periodicity in their \textit{CoRoT} lcs 
(Mon-296, Mon-1308), a structure explaining this can be a fixed warp 
in the inner section of the disk that is repeated periodically along 
the line of sight. The lc for Mon-314 has
a low-probability periodicity in the 2011 epoch of observation and is
non periodic in the 2008 epoch which indicates that the physical
mechanism shaping the lc is not long-lived, thus, a warp may or
may not be the cause to explain the 2011 epoch lc. For Mon-433 in
any of the two epochs the lc is not periodic, thus, the shape cannot
be described with a stable warp. However, looking the lc, there is 
a clear sequence of peaks and valleys, such that the occulting 
structures are locally periodic (they are moving at the keplerian
velocity according to their location) but the dominant one is not
always the same, meaning that the period is changing. Also from the
lc we can conclude, that the timescale of the variability is a few 
days indicating that the structures are located close to the inner
edge of the disk. The shaping of this region is given by the interaction 
of the disk with the stellar magnetic field lines \citep{2013MNRAS.430..699,2020A&A...643..A157}. 

\begin{table}[!t]\centering
  \setlength{\tabnotewidth}{0.5\columnwidth}
  \tablecols{8}
  \setlength{\tabcolsep}{2.0\tabcolsep}
  \caption{CoRoT light curves periods for the 2008 and 2011 campaign }
 \begin{tabular}{ccc}
    \toprule
    Object & $\rm{P}$(days), 2008 campaign & $\rm{P}$(days), 2011 campaign  \\
    \midrule
     Mon-296  & $2.51$ & $3.91$  \\
     Mon-314  & -----  & $3.38$  \\
     Mon-433  & -----  & -----   \\
     Mon-1308 & $6.45$ & $6.68$  \\
    \bottomrule
  \end{tabular}
  \label{tab:periods}
\end{table}

Another aspect to be pointed out is the resemblance or not of the
\textit{CoRoT} and \textit{Spitzer} lcs: for Mon-296 both lcs are completely different,
for Mon-314 and Mon-433 there is a resemblance between both lcs and
finally for Mon-1308, the resemblance is remarkable. A high 
resemblance between the lcs suggest a strong connection between the 
mechanism explaining them.

The observational values that we aim to model are the amplitudes
for the optical ($\Delta mag_{obs}$) and the IR lcs 
($\Delta mag_\mathrm{IR,obs}$). These values come from data of the \textit{CoRoT} and 
\textit{Spitzer} telescopes extracted by \citet{2015A&A...577..A11} and included in 
Table~\ref{tab:amplitude_mag_obs}. $\Delta mag_\mathrm{obs}$ comes from the \textit{CoRoT} Telescope and $\Delta mag_\mathrm{IR,obs}$ comes from the \textit{Spitzer} telescope. The \textit{Spitzer} photometric data includes values for wavelengths at $3.6$ and $4.5\mu$m. Because the behavior is similar
in both wavelengths we choose $3.6\mu$m for the analysis.

\begin{table}[!t]\centering
  \setlength{\tabnotewidth}{0.5\columnwidth}
  \tablecols{3}
  \setlength{\tabcolsep}{2.8\tabcolsep}
  \caption{Observed amplitudes for the lcs}
 \begin{tabular}{ccc}
    \toprule
    Object & $\Delta mag_\mathrm{obs}^{\,\,\,a}$ & $\Delta mag_\mathrm{IR,obs}^{\,\,\,b}$ \\
    \midrule
     Mon-296  & $0.2-0.7$  & $<0.1$       \\
     Mon-314  & $0.1-0.15$ & $0.1-0.15$   \\
     Mon-433  & $0.2-0.4$  & $0.2-0.4$    \\
     Mon-1308 & $0.2-0.4$  & $0.2-0.4$    \\     
    \bottomrule
    \tabnotetext{a}{Values from the \textit{CoRoT} Telescope}
    \tabnotetext{b}{Values from the \textit{Spitzer} Telescope}
  \end{tabular}
  \label{tab:amplitude_mag_obs}
\end{table}

\section{Modeling}
\label{sec:modeling}

For the modeling of the optical lc for each object, we modify the 
code used in \citet{2019A&A...625..A45} to include the extinction by dust 
in a disk hole where the contribution of an optically thick region is neglected. 
Besides, we include the emission of the dust located in the hole in order to
consistently calculate the modeled IR lc. As mentioned, the lcs are completely 
shaped by the spatial dust distribution which is given by the gas density, 
the gas (and dust) is inwards limited by the magnetospheric radius, $R_\mathrm{mag}$,
which is assumed to equal the keplerian radius ($R_\mathrm{k}$) at the 
period extracted from the optical lcs and outwards limited by the 
hole radius ($R_\mathrm{H}$) given in \citet{2019A&A...629..A67}. 
The optically thin material is distributed according to the gas density 
$\rho$ which depends on the azimuthal angle $\phi$ and the vertical
coordinate $z$ and is given by

\begin{equation}
 \rho=A\cos({\phi-\phi_{0}\over 2})e^{-z/{H_H}}(1-\beta 
 {R\over R_{H}}), 
 \label{eq:rho}
\end{equation}

where $\phi=\phi_{0}$ is the location where the maximum density is found and
the factor $1/2$ allows to have only one maximum in the $\phi$ range. 
The value for $\phi_{0}$ is choosen such that at $phase=0.5$ (center of the 
plotted lc) the maximum in density (and the minimum in optical flux) is along 
a line of sight towards the star.
This density peak is responsible to periodically occult the star as it
is required to interpret the optical lc. The argument inside the exponential 
simply model a natural concentration tendency towards the midplane of the 
disk. We do not assume that the hole material has reached a vertical hydrostatic 
equilibrium. The scale height $H_\mathrm{H}$ represents a width of the accreting 
stream in the hole which we fix as $H_\mathrm{H}=0.1R_\mathrm{H}$ a value typical of
the disk scale height at the location where the material "falls" to the hole from a
stationary disk in vertical hydrostatic equilibrium. Because the velocity
in the stream is much larger than the accreting velocity in the disk then
the density in the hole is lower as it is required to get an optically
thin environment. As the timescales between disk and hole are different,
it is safe to assume that in the hole, the vertical equilibrium is not reached. 
In any case, the exact shape of the functional form of $\rho$ lacks of relevance 
for the support of the main conclusions presented later in this work
because the amplitude of the optical lc is mainly given by the density maximum
whose order of magnitude is given by $\dot{M}$ and not by the
functional form of $\rho$.

The free parameter $\beta$ models the radial concentration of dust/gas: 
$\beta=0$ corresponds to a homogeneous distribution and $\beta=1$ indicates
that the material is concentrated towards the inner edge of the hole. Note
that this latter case is close to what one expects for a structure moving at
the observed periodicity ($3-10\,\,days$) for the sample of objects, where 
$R_{k}$ is located in the inner region. We do not include a detailed
analysis of dust sublimation. Even for the $\beta=1$ case, most of the region
responsible to shape the lcs is beyond the magnetospheric radius which is
the lower limit for the grid used.
The constant coefficient $A$ is calculated assuming two
facts; the first one is that all the material incorporating to the hole 
(at $R_{H}$) arrives at the star in the free-fall time given by $t_\mathrm{ff}={R_\mathrm{H}\over v_\mathrm{ff}}$ where

\begin{equation}
 v_\mathrm{ff}=\sqrt{2GM_{\star}\over R_{\star}}\sqrt{1-{R_{\star}\over R_\mathrm{H}}}
\end{equation}

is the free-fall velocity. The second fact is that $\dot{M}$ resulting of 
this process is equal to the observed value, $\dot{M}_\mathrm{obs}$. 

Note that the dust located in the gas distributed as in equation~\ref{eq:rho} 
is the one responsible to shape the IR lc. In a fully optically
thin stellar surroundings, the IR photometric variability is small because
the dust grain emission is not extincted or blocked. However, in the system 
configuration there are grains occulted behind the star. For Mon-296 whose 
hole is small, the fractional area representing this occultation 
($f_\mathrm{A}={\pi R_{\star}^2\over \pi R_\mathrm{H}^2 \cos i}$)
amounts to $0.009$ such that the blocking of this fraction of the emission corresponds
to $\Delta mag_\mathrm{IR}\sim -2.5\log10 (1-f_\mathrm{A})\sim 0.01$ which is of the order of magnitude of 
$\Delta mag_\mathrm{IR,obs}$ as can be seen in Table~\ref{tab:amplitude_mag_obs}. For
Mon-314,433 and 1308 the variability coming from this geometrical occultation
is $3$ to $4$ orders of magnitude smaller than for Mon-296. This means that
for the $3$ objects set, the dust extinction is relevant to search a 
physical configuration prone to explain $\Delta mag_\mathrm{IR,obs}$.
    
We assume that all the material in the hole is moving at the same rotational
velocity, however this is not true. The expected orbit for each 
accreting particle is a spiral, because at $R_\mathrm{H}$ and $R_\mathrm{k}$ the orbital 
Keplerian period in a circular trajectory are $\sim4000$days and $6.45$days, 
respectively which can be compared to $t_\mathrm{ff}=89$days where the parameters for Mon-1308
are used.
In other words, during the free-fall, the particle gives several orbits around
the star until the arrival to the surface. 
The assumption of a dynamical model will locally 
change the dust distribution but to fit $\Delta mag_\mathrm{obs}$ the most important 
parameter is the maximum of the surface density along the line of sight. The maximum is calculated using $\rho$, and this latter will not change apreciably using a detailed model.

\section{Results}
\label{sec:results}

The fitting of $\Delta mag_\mathrm{obs}$ and $\Delta mag_\mathrm{IR,obs}$ is done using two free parameters: $\beta$ and the dust to gas ratio $\zeta$. The latter is 
parameterized by $\alpha$ where 
$\zeta=\alpha\zeta_\mathrm{typ}$ and $\zeta_\mathrm{typ}=0.01$ is assumed as typical for protoplanetary disks. According to the model, the value $\beta=1$ corresponds 
to the configuration with the largest concentration of material close to the star
which as mentioned in Section~\ref{sec:modeling} is a physically expected 
configuration. For this reason, the fiducial model is defined by $\alpha=1$ and 
$\beta=1$.

For each object, $\dot{M}_\mathrm{obs}$ is given as in table~\ref{tab:observed_param} 
and $H_{H}$ is fixed to $H_\mathrm{H}=0.1R_\mathrm{H}$. We note that there is a 
degeneracy between $\dot{M}$ and $\zeta$ because if both parameters 
increase/decrease then the amount of dust increases/decreases keeping the
shape of the spatial distribution of material. We decided to fix $\dot{M}$ at 
$\dot{M}_\mathrm{obs}$ because is consistent with the estimates based on observations,
and therefore interpret the model according to changes in $\zeta$ ($\alpha$). 
The value for $\Delta mag_\mathrm{IR}$ is assigned to the maximum magnitude change 
for the $3.6\mu$m \textit{Spitzer} band. Note that the behaviour for the $4.5\mu$m band 
is similar as can be seen in the \textit{Spitzer} observations shown in 
\citet{2015A&A...577..A11} for the set of objects studied here. 

We find models consistent with the observed range of $\Delta mag_\mathrm{obs}$ and 
$\Delta mag_\mathrm{IR,obs}$ using a grid of models in the ranges of $\alpha$ and 
$\beta$ given by $[0.1,10]$ and $[0,1]$, respectively. 
In Table~\ref{tab:models}, we present the parameters for the representative consistent models found in \S~\ref{sec:Mon-1308},\S~\ref{sec:Mon-433},\S~\ref{sec:Mon-314}, and
\S~\ref{sec:Mon-296}. The first column corresponds to the object name, 
the second column is $\alpha$, and the third column is $\beta$.
The next three columns are associated to the optical lc; the minimum flux of the star, 
namely $F_\mathrm{min,opt}$, the maximum flux of the star, namely $F_\mathrm{max,opt}$ and $\Delta mag$.
The final three columns are associated to the IR lc; the minimum total flux of the star plus disk, namely $F_\mathrm{min,IR}$, the maximum total flux of star plus disk, namely $F_\mathrm{max,IR}$ and $\Delta mag_\mathrm{IR}$. All the fluxes are given in units of $10^{-12}\,\mathrm{erg\,cm^{-2}\,s^{-1}}$.

\begin{table}[!t]\centering
  \setlength{\tabnotewidth}{0.5\columnwidth}
  \tablecols{9}
  \setlength{\tabcolsep}{1.8\tabcolsep}
  \caption{Parameters for representative models}
 \begin{tabular}{ccccccccc}
    \toprule
    Object & $\alpha$ & $\beta$  & $F_\mathrm{min,opt}^{a}$ & $F_\mathrm{max,opt}^{a}$ & $\Delta mag$ & $F_\mathrm{min,IR}^{a}$ & $F_\mathrm{max,IR}^{a}$ & $\Delta mag_\mathrm{IR}$ \\
    \midrule
     Mon-296 & 1  & 1   & $47.7$ & $57.6$ & $0.226$ & $5.30$ & $5.35$ & $1.05\times 10^{-2}$  \\
     Mon-296 & 10 & 0.5 & $30.3$ & $55.0$ & $0.719$ & $8.01$ & $8.33$ & $4.37\times 10^{-2}$  \\
     Mon-314 & 0.5 & 0.7 & $3.267$ & $3.60$ & $0.107$ &  $1.610$ & $1.619$ & $5.68\times 10^{-3}$ \\
     Mon-314 & 0.7 & 1 & $3.265$ & $3.60$  & $0.107$ & $1.612$ & $1.620$ & $5.58\times 10^{-3}$ \\
     Mon-433& --- & --- & --- & --- & --- & --- & --- & ---     \\
     Mon-433& --- & --- & --- & --- & --- & --- & --- & ---     \\
     Mon-1308& 5 & 1 & $10.8$ & $14.1$ & 0.284 & $2.94800$ & $2.94812$ & $4.29\times 10^{-5}$     \\
     Mon-1308& 7 & 1 & $9.79$ & $14.1$ & 0.398 & $2.94803$ & $2.94819$ & $5.98\times 10^{-5}$     \\
    \bottomrule
    \tabnotetext{a}{All the fluxes are given in units of $10^{-12}\,\mathrm{erg\,cm^{-2}\,s^{-1}}$.}
  \end{tabular}
  \label{tab:models}
\end{table}

\subsection{Mon-1308}
\label{sec:Mon-1308}

For both, $\Delta mag_\mathrm{obs}$ and $\Delta mag_\mathrm{IR,obs}$, the range is $[0.2,0.4]$. A fit for $\Delta mag_\mathrm{obs}$ is found when $\alpha=5$ (with $0.8<\beta<1$), $\alpha=7$ (with $0<\beta<1$) and $\alpha=10$ (with $0<\beta<0.8$). Due to the grid of models, we run models for $\alpha=5,7$ and $10$ but not intermediate values;
however another models with other values of $\alpha$ between $5$ and $10$ also
are inside the range for $\Delta mag_\mathrm{obs}$. For these
models $\Delta mag_\mathrm{IR}<10^{-4}$ which is $3$ orders of magnitude lower than
required for a reasonable fit.
The parameters for the models with $\beta=1$ are shown in Table~\ref{tab:models}.
 
In order to favor these models, either the dust to gas mass ratio should be 
higher than expected in a typical disk, or $\dot{M}$ should be outside the
observational estimates. In any case, even if this can be achieved, only 
$\Delta mag_\mathrm{obs}$ can be explained leaving the fitting of $\Delta mag_\mathrm{IR,obs}$ to another physical mechanism. 

It is not expected IR photometric variability in 
an optically thin stationary system because all the emitting material contributes to the IR lc. However, this is not true if there are an extinction and an occultation mechanism; namely the dust extincts and the star occults the dust behind it. Including both mechanisms, the modeling of the variability in the IR is small compared to the one associated to the optical. The previous result and the fact that for Mon-1308, $\Delta mag_\mathrm{obs}\sim \Delta mag_\mathrm{IR,obs}$, explains our inability to consistently model IR and optical lcs using optically thin material in the hole of a TD.

The last analysis lead us to conclude that instead of a TD, a possible 
configuration to simultaneously explain both the optical and the IR lcs 
is a PTD. In spite of the classification of \citet{2009AJ...138..1116} where Mon-1308 is catalogued as a TD due to the amount of dust in the inner disk, we try to interpret this object as a PTD. For the modeling of Mon-1308 in 
\citet{2019A&A...625..A45} is required an optically thick warp located at 
$R_\mathrm{mag}$ meaning that the remaining of the emission should be associated to 
the disk outwards the warp. \citet{2019A&A...625..A45} do not explicitly 
characterize the shape of the disk but in accordance with the analysis 
given here, the thick warp plays the role of the inner disk of the PTD. 

Taken together, the previous discussion should be connected to the  
SED fitting code used in \citet{2019A&A...629..A67} which only includes the
dust emission outside an empty hole. In the model presented here, the dust emission 
should come from an optically thin hole and the outer disk. We point out 
that in our model the variability is completely associated to the dust
distribution inside the hole. In the case of a PTD,
the observed emission should be interpreted as coming from both an inner, 
and outer disks, and the dusty gap. A follow up goal can be to find a SED fitting 
consistent with the model described in \citet{2019A&A...625..A45} where
the innermost structure is a sublimation wall covering a small radial range.

The optical flux coming from our TD model spans between $F_\mathrm{min,opt}$ and $F_\mathrm{max,opt}$, values given in Table~\ref{tab:models}. These values are consistent with the SED fitting in \citet{2019A&A...625..A45}.
For Mon-1308, the model presented in \citet{2019A&A...625..A45} shows that
the flux at $4.5\mu$m coming from the full disk, namely $F_\mathrm{disk,4.5}$ is 
similar to the flux at $4.5\mu$m coming from the star, namely $F_{\star,4.5}$
which is around half the observed flux, namely $F_\mathrm{obs,4.5}$. 
In Table~\ref{tab:models} the range of fluxes modeled at $3.6\mu$m are presented,
spanning between the value $F_\mathrm{min,IR}$ and $F_\mathrm{max,IR}$. These values are
consistent with $F_\mathrm{disk,4.5}+F_{\star,4.5}$. In the 
model by \citet{2019A&A...629..A67}, $F_{\star,4.5}\sim F_\mathrm{obs,4.5}$,
because there is a negligible contribution of any part of the disk. 
We note that in their modeling $T_{\star}$ (free parameter) is 
$1091\,\mathrm{K}$ larger than the values estimated in 
\citet{2014A&A...570..A82}, allowing us to suggest that a new model 
including an inner disk and a star with a lower $T_{\star}$ as the 
main contributors in the near-IR is a reasonable aim. Besides, note that in
\citet{2015A&A...577..A11}, $F_\mathrm{disk,4.5}/F_{\star,4.5}=0.7$
meaning that there is a clear contribution of material around the star. This suggest
that PTD or full disks are configurations prone to explain this excess.

In order to pursue this further, we find a new synthetic SED using 
the Python-based fitting code Sedfitter \citep{2017A&A...600..A11} based on the 
3D dust continuum radiative transfer code Hyperion, an open-source parallelized 
three-dimensional dust continuum radiative transfer code by \citet{2011A&A...536..A79}, 
which was used for the modeling in \citet{2019A&A...629..A67}.
This code is composed of modular sets with components that can include a stellar 
photosphere, a disk, an envelope, and ambipolar cavities. To model Mon-1308, we used 
two sets of models. Model 1 is composed of a stellar photosphere and a passive disk, 
model 2 - stellar photosphere, passive disk, and a possible inner hole. The Hyperion 
SED model includes only a passive disk such that does not consider the disk heating due to
accretion. The input 
parameters of the Hyperion Sedfitter are a range of Av, the distance from the Sun, 
the fluxes, and its uncertainties. The input fluxes uses UBVRcIc optical photometry from 
\citet{2002AJ...123..1528}, near-IR photometry JHKs from 2MASS, \textit{IRAC} 
\citep{2004ApJS...154..10} and MIPS \citep{2004ApJS...154..25} magnitudes from 
\textit{Spitzer} satellite, and \textit{WISE} observations at 3.4, 4.6, 12.0, and 
22$\mu$m \citep{2010AJ...140..1868}. We used the distance estimated from parallax data 
obtained from the \textit{Gaia} second release \citep{2016A&A...595..A1,2018A&A...616..A1}.

\begin{figure*}[!t]
  \includegraphics[width=0.45\linewidth,height=3cm]{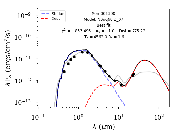}%
  \hfill
  \includegraphics[width=0.45\linewidth,height=3cm]{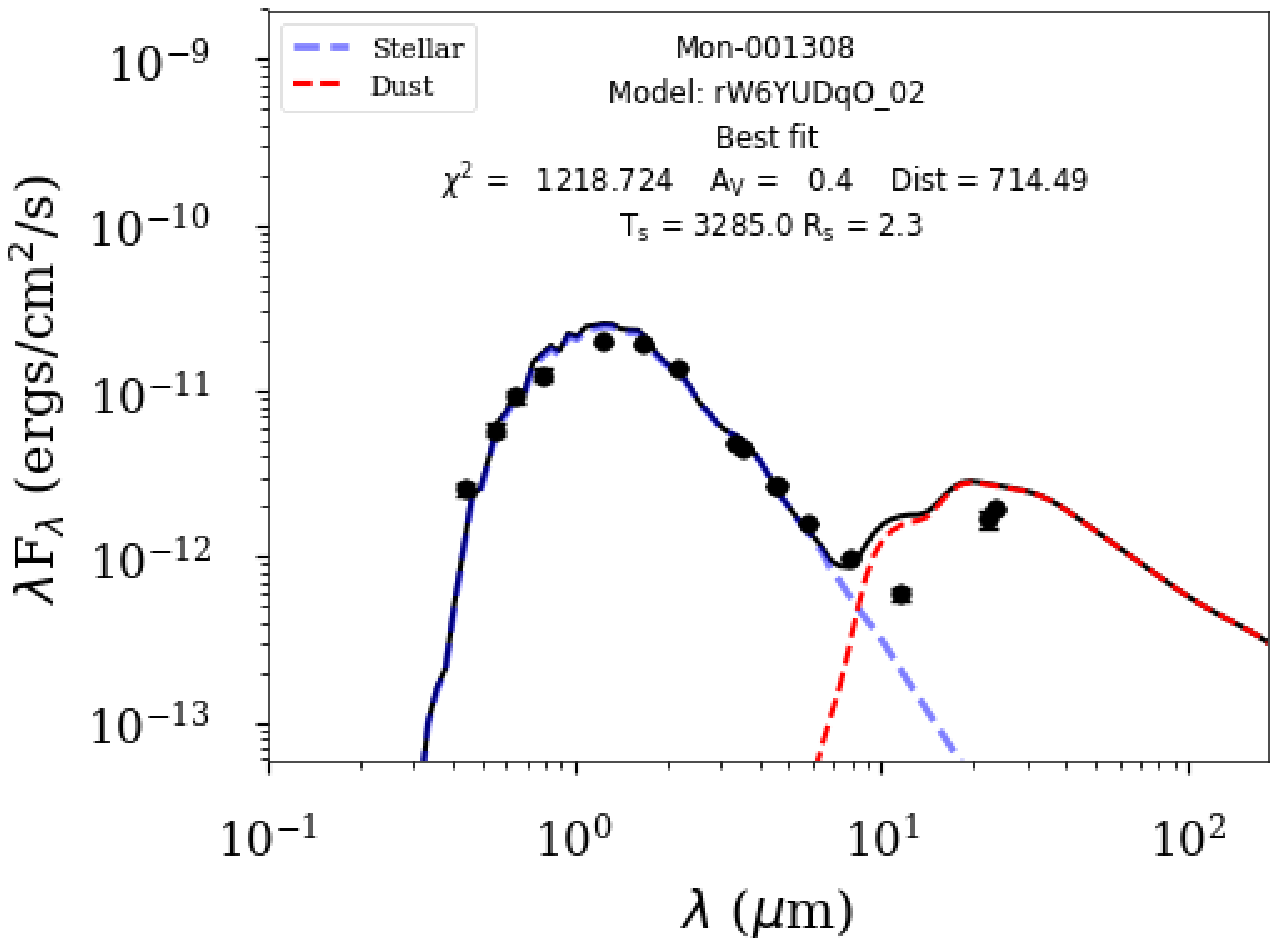}
  \caption{Two new SED fits for Mon-1308 using the 3D dust continuum radiative transfer code Hyperion. The SED in the left corresponds to the best fit using a stellar photosphere and a full disk. The SED in the right corresponds to the best fit using a stellar photosphere and a disk with an inner hole (TD). The plot includes the values for the fitted parameters. The emission comes from the star (blue line), and a full-disk or TD (red line). The total flux is shown as a black line.}
  \label{fig:Mon-1308_SEDfit}
\end{figure*}

This code has not a setup for a PTD, thus, we are not able to test this scenario.
For the new fit, we remove the SDSS magnitudes because we realized that they are not as trustworthy compared to UBVRI data. The magnitudes are input parameters in the code. Besides, using a different $A_\mathrm{v}$ range from \citet{2019A&A...629..A67}; we find a new fit for the data with a full disk model with more consistent $T_{\star}$ and inclination $i$ values. Instead of $A_\mathrm{v}=0.2$, $T_{\star}=5011$K and $i=86.4^{\circ}$, the new values are $A_\mathrm{v}=1$, $T_{\star}=4562$K and $i=60.99^{\circ}$. We try to fit the observational data using a TD but the $i$ obtained is $i=18.8^{\circ}$ value not consistent with the variability observed in the lcs which requires dusty structures that intermittently block the stellar radiation only present in a high-i configuration. As a secondary argument, against the TD model $\chi^2=1218.724$ compared to $\chi^2=853.496$ for the full-disk case. Both new fits are shown in Figure~\ref{fig:Mon-1308_SEDfit}.

Summarizing, there are three important facts. The first one is that our modeling of the lcs is unsuccessful, thus, the presence of dust in the inner hole (a TD) is not consistent with this part of the observations. The second fact is that the PTD scenario cannot be tested. The third one is that the new fitting of the SED suggest that a full disk in Mon-1308 is reasonable.
Our conclusion is that a full disk could be a possible option for this object.

\subsection{Mon-433}
\label{sec:Mon-433}

For both, $\Delta mag_\mathrm{obs}$ and $\Delta mag_\mathrm{IR,obs}$, the range is $[0.2,0.4]$. A fit for $\Delta mag_\mathrm{obs}$ is not found for the ranges of $\alpha$ and $\beta$ studied ($0.1<\alpha<10$,$0<\beta<1$). For these models $\Delta mag<0.1$ which is not consistent with the observations. We require an increment of more than 
one order of magnitude in $\dot{M}$ or on the dust to gas ratio to explain 
$\Delta mag_\mathrm{obs}$ but in any case we are not close to interpret 
$\Delta mag_\mathrm{IR,obs}$. The larger observational estimate of $\dot{M}$ is only
$1.23$ times the value used for the modeling, thus, the evidence leads us to
conclude that alone, the optically thin material in the hole is not enough to
interpret the lcs in the optical and in the IR. 
 
 In Table~\ref{tab:periods} we show that the periodicity analysis
done in \citet{2014AJ...147..82} indicates that there is not a clearly defined period
for the CoRoT lc of Mon-433. This is summarized in \citet{2015A&A...577..A11} where 
they catalogued the CoRoT light curve in 2011 as aperiodic. 
However, in a timescale of around 5 to 10 days, there is a sequence of peaks and valleys that 
indicates that underneath it there is some periodic physical structure 
as the one presented in this work. Around this main periodic structure there is another set 
of minor structures located at different places with different periods that shape the 
observed light curve. This piece of evidence lead us to a physically reasonable testing 
of this model. As in Mon-1308, for Mon-433 a PTD includes a structure responsible 
to add another component for the extinction in order to explain the observations. 
We are unable to test this using the SEDfitter code based on Hyperion because it has not 
a setup for a PTD, just full disks and TDs.

\subsection{Mon-314}
\label{sec:Mon-314}

For both, $\Delta mag_\mathrm{obs}$ and $\Delta mag_\mathrm{IR,obs}$, the range is $[0.1,0.15]$. A fit for $\Delta mag_\mathrm{obs}$ is found when $\alpha=0.3$ (and $0<\beta<0.4$), $\alpha=0.5$ (and $0.1<\beta<0.7$) and $\alpha=0.7$ (and $0.4<\beta<1$). 
Due to the grid of models, we run models for $\alpha=0.3,0.5$ and $0.7$ but not 
intermediate values; however another models with other values of $\alpha$ between 
$0.3$ and $0.7$ also are inside the range for $\Delta mag_\mathrm{obs}$. For 
these models $\Delta mag_\mathrm{IR}<10^{-2}$ which is one order of magnitude lower than
required for a reasonable fit. The optical flux coming from the model spans between 
$F_\mathrm{min,opt}$ and $F_\mathrm{max,opt}$, values given in Table~\ref{tab:models}. 
In Table~\ref{tab:models} the range of fluxes modeled at $3.6\mu$m are presented,
spanning between the value $F_\mathrm{min,IR}$ and $F_\mathrm{max,IR}$. These values are
consistent with the modeling in \citet{2019A&A...629..A67}.
Note that the values for $\alpha$ implies that the dust-to-gas ratio is lower
than typical for protoplanetary disks. However, we think that the models are 
physically reasonable to explain $\Delta mag_\mathrm{obs}$ but its inability to 
explain $\Delta mag_\mathrm{IR,obs}$ lead us to look for another configuration. 

As the objects Mon-1308 and Mon-433, Mon-314 also satisfy that 
$\Delta mag\sim \Delta mag_\mathrm{IR}$ such that also all the analysis developed
in \S~\ref{sec:Mon-1308} are valid here. Look in Table~\ref{tab:models}
for some parameters coming from the modeling of two
cases explaining $\Delta mag_\mathrm{obs}$. 

Note that \citet{2015A&A...577..A11} do not find a stable period in the lc,
however, it is clear that the physical mechanism repeats itselfs because
within a temporal range, a sequence of peaks and valleys are clearly seen
in the lc. Thus, it is valid to try a periodical model to fit the 
photometric data. As for Mon-433, a likely model for Mon-314
is a PTD.

\subsection{Mon-296}
\label{sec:Mon-296}

For $\Delta mag_\mathrm{obs}$ and $\Delta mag_\mathrm{IR,obs}$, the range is $[0.2,0.7]$ and $[0,0.1]$, respectively. A fit for $\Delta mag_\mathrm{obs}$ is found 
when $\alpha=0.7$ (and $0<\beta<0.3$), $\alpha=1$ (and $0<\beta<1$), $\alpha=3$ 
(and $0<\beta<1$), $\alpha=5$ (and $0<\beta<0.8$), $\alpha=7$ (and $0<\beta<0.7$) and $\alpha=10$ (and $0<\beta<0.5$). Due to the grid of models, we run models for 
$\alpha=0.7,1,3,5,7$ and $10$ but not intermediate values; however another models with 
other values of $\alpha$ between $0.7$ and $10$ also are inside the range for 
$\Delta mag_\mathrm{obs}$.
For these models $\Delta mag_\mathrm{IR}$ is between $1.05\times 10^{-2}$ and $4.37\times 10^{-2}$ which is within an order of magnitude lower than required for a reasonable fit. The optical flux coming from the model spans between $F_\mathrm{min,opt}$ and 
$F_\mathrm{max,opt}$, values given in Table~\ref{tab:models}. These values are consistent with the SED fitting in \citet{2019A&A...629..A67}.

\begin{figure}[!t]
  \includegraphics[width=\columnwidth]{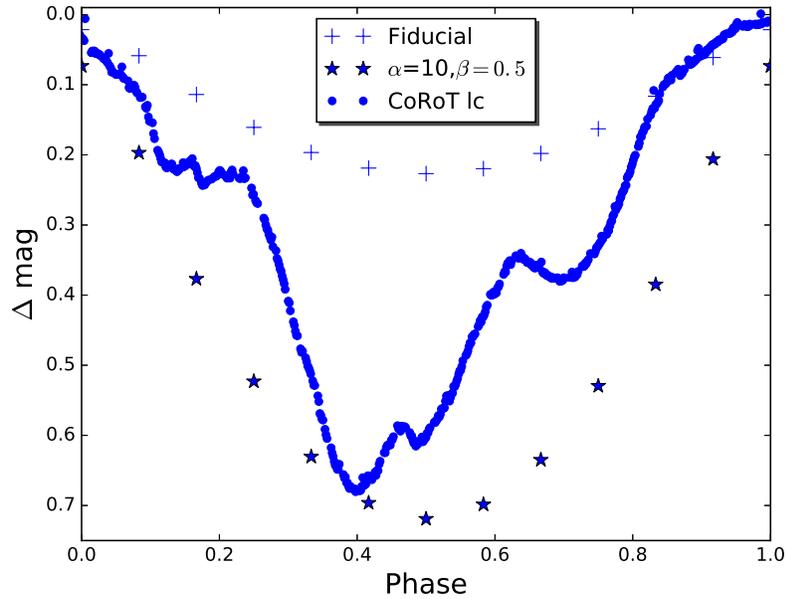}
  \caption{Two models for the optical lc of Mon-296. The first model is shown with
plus signs, corresponding to the fiducial: $\alpha=1$ and $\beta=1$. The second
model is represented with stars, corresponding to $\alpha=10$ and $\beta=0.5$. The points
show a representative section of the CoRoT lc in the 2011 campaign.}
  \label{fig:Mon-296_Opt}
\end{figure}

\begin{figure}[!t]
  \includegraphics[width=\columnwidth]{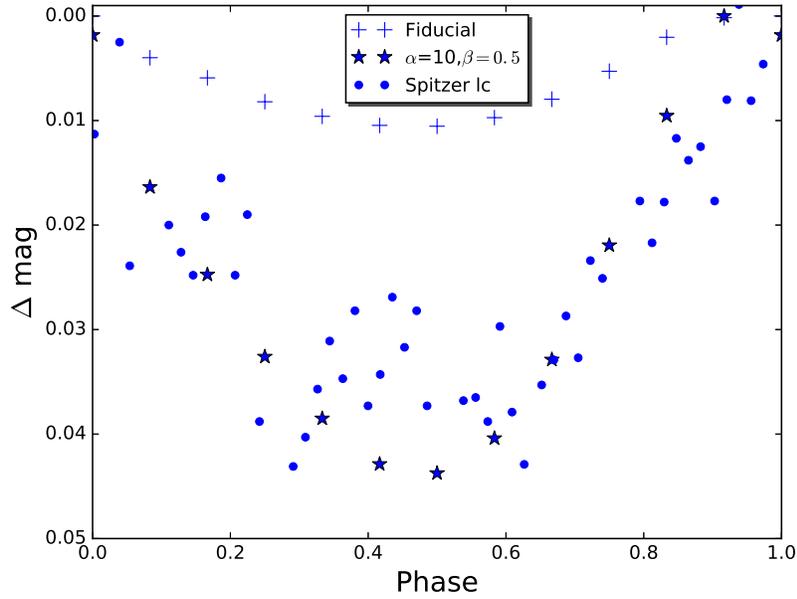}
  \caption{Two models for the IR lc of Mon-296. The first model is shown with
plus signs, corresponding to the fiducial: $\alpha=1$ and $\beta=1$. The second
model is represented with stars, corresponding to $\alpha=10$ and $\beta=0.5$. The points
show a representative section of the Spitzer lc in the 2011 campaign.}
  \label{fig:Mon-296_IR}
\end{figure}

For Mon-314, Mon-433, and Mon-1308, $\Delta mag_\mathrm{IR}$ is among two and three orders of magnitude lower than $\Delta mag_\mathrm{IR,obs}$. For the fiducial model for Mon-296, $\Delta mag=0.226$ and $\Delta mag_\mathrm{IR}=0.01$. This is the only system where the fiducial model is consistent with $\Delta mag_\mathrm{obs}$. For a $10$ times more massive hole ($\alpha=10$) with $\beta=0.5$, both values increase to: $\Delta mag=0.719$ and 
$\Delta mag_\mathrm{IR}=0.0437$, the last value is consistent with the observed IR lc. The modeled lcs corresponding to the previous cases are presented in 
Figure~\ref{fig:Mon-296_Opt} for the optical and in Figure~\ref{fig:Mon-296_IR} 
for the IR. Also in the figures we include a section of the CoRoT and 
Spitzer lcs for campaign 2011.
 An increase of the dust abundance and/or $\dot{M}$ is required to 
explain the lcs within the framework of this modeling. Another possibility is
that the material in the hole is not completely thin but coexists with partial 
or completely optically thick structures like streams that connect the outer
disk with the star.  

In Table~\ref{tab:models} the range of the total flux modeled at $3.6\mu$m for the fiducial and the massive model are presented, spanning between the value $F_\mathrm{min,IR}$ and $F_\mathrm{max,IR}$. In the models of the optical, the total flux range spans between $F_\mathrm{min,opt}$ and $F_\mathrm{max,opt}$. These values are consistent with the modeling in \citet{2019A&A...629..A67}.
If we extract the contribution of the stellar flux in the optical and in the IR then $F_{\star}\sim 5\times 10^{-11}$ and $F_\mathrm{\star,IR}\sim 5\times 10^{-12}$ which are consistent with the SED presented in \citet{2019A&A...629..A67}. In the fiducial case, the IR flux of 
the material in the hole is $F_\mathrm{IR}\sim 5\times 10^{-13}$, which is one order of magnitude lower than the flux associated to the optically thick disk required 
for the modeling using the SED fitting code based in Hyperion \citep{2019A&A...629..A67}. 
However, the hole material for the massive model produces a flux given by 
$F_\mathrm{IR}\sim 3.29\times 10^{-12}$ where $F_\mathrm{IR}/F_{\star,IR}=0.7$ just a factor of two lower than the observational estimate of $1.6$ for $4.5\mu$m in
\citet{2015A&A...577..A11}. The value for $F_\mathrm{IR}$ is estimated as the excess above the photospheric template used to calculate $F_\mathrm{\star,IR}$. This means that in the latter case, the flux of the material in the hole does notoriously affect the SED fitting, but on the other hand is necessary to explain both the lcs in the optical and in the IR. 

An estimate of the surface density required to calculate the extinction is 
done using the constant parameter A in Equation~\ref{eq:rho} such that the largest 
value corresponds to Mon-296. This explains the existence of a lot of models consistent with $\Delta mag_\mathrm{obs}$. Also, from our set of objects, this is the only one
satisfying $\Delta mag >> \Delta mag_\mathrm{IR}$ resulting that the emission and
occultation caused by a small optically thin hole is consistent either with
the SED as shown in \citet{2019A&A...629..A67} and the \textit{CoRoT} and \textit{Spitzer}
lcs presented in \citet{2015A&A...577..A11}. Note that the optically thick disk
outwards the hole can produce occultations for inclinations larger than 
$atan(R_\mathrm{H}/H_\mathrm{H})=84.29^\circ$ meaning that for this object the occultation 
structure is located inside the hole. For the other objects, occultations caused
by the outer edge of the hole are not relevant looking at the evidence in the lcs,
namely that the variability timescale is associated to their inner regions.

\section{Discussion}
\label{sec:discussion}

For Mon-1308, it is required the presence of a small optically thick
inner disk in order to simultaneously explain the optical and IR lcs, 
because in this case both lcs has a strong resemblance such 
that the physical region shaping both is the same.
As noted in \S~\ref{sec:Mon-1308}, we are able to explain the 
optical lc with the material concentrated at the inner edge of the disk. 
However, in order to fit the IR lc, an increment in the amount of dust for several orders of magnitude is required, which is not physically correct.

Because it is not possible to find a physically correct modeling for the
optical and IR lcs for Mon-314 and Mon-433, then we suggest that these two systems require a small optically thick
disk in the innermost regions responsible to shape both lcs. This is 
consistent with the LkCa 15 TD because it has a $50$AU wide cavity 
\citep{2011ApJ...742..L5} but also photometric
variability with a very small period which locates a disk-like
structure responsible to it very close to the star 
\citep{2018A&A...620..A195}.

For Mon-296, $R_\mathrm{k}$ consistent with the stellar rotational period $P$ is 
less than $R_\mathrm{H}$, thus the material responsible to shape the periodic 
optical lc is inside the optically thin environment. For this object, 
$R_\mathrm{k}=0.054$AU and $R_\mathrm{H}=0.12$AU, such that $R_\mathrm{ovr}=0.086$AU is inside this range.  
$R_\mathrm{ovr}$ is the location of the outer vertical resonance which comes from
the analysis of the propagation of small-amplitude waves in the disk, here the
study of out-of-plane gravity modes implies the excitation at this radius of
bending waves. This analysis comes from the linearization of the equations
of motion for the fluid in the disk where the external force is calculated with the
stellar magnetic field which is moving at the
stellar rotational velocity \citep{2013MNRAS.430..699}. The bending wave
rotates at the stellar rotational velocity, thus is a structure prone to 
explain periodicity at the period corresponding to this velocity.
Note that a value of $\beta=0.9$ means 
that $66\%$ of the material is moving at the velocity required to have 
a periodic feature with the period $P$, however, the model explaining
$\Delta mag_\mathrm{obs}$ corresponds to $\beta=0.5$ resulting that not all
the dust is distributed in a region where the material is rotating
with the period $P$. It is important to mention that the modeled 
optical and IR lcs evolve with the same periodicity but for Mon-296 
there is not a clear resemblance between the observed lcs. This leads 
to an interpretation where the actual dust distribution is not completely
described by the density $\rho$ given in equation~\ref{eq:rho} such 
that this $\rho$ works as the backbone of the actual structure where
the multiple features expected in a highly dynamical environment
near the stellar magnetosphere are important to explain the details
of the lcs.

Optically thin dust emission does not depend on the lc phase, thus, 
the low amplitude magnitude shown in the \textit{Spitzer} lc for Mon-296 is 
consistent with this fact; note that at many times the IR magnitude 
is constant. The small value for $R_{H}$ means that an optically 
thick disk is close enough to the star to have an adequate temperature 
to contribute in the IR as can be seen in \citet{2019A&A...629..A67}. 
This is important because the physical configuration for the model presented
here results in similar shapes for the optical and IR lcs which is the
opposite compared with the observations. Thus, we can argue that an
optically thick disk flux contribution changing with phase is relevant 
as a second mechanism to fully interpret the IR variability.

\section{Conclusions}
\label{sec:conclusions}

1.- For Mon-314 and Mon-1308, $\Delta mag\sim \Delta mag_\mathrm{IR}$ 
but the model predicts $\Delta mag >> \Delta mag_\mathrm{IR}$, thus, we are
able to model $\Delta mag_\mathrm{obs}$ but not consistently model 
$\Delta mag_\mathrm{IR,obs}$. Using the grid of models defined within the range
$0.1<\alpha<10$ and $0<\beta<1$, we cannot find a consistent model for
Mon-433. Mon-296 is the only system where 
$\Delta mag >> \Delta mag_\mathrm{IR}$, such that this is the only object
that can be modeled using the optically thin material inside the disk. 

2.- The density in a small optically thin hole ($\sim$tenths of AU) is large enough to explain typical amplitudes ($\sim$tenths) of the optical lc produced by dust extinction of the stellar spectrum. Using the observed 
stellar and disk parameters, Mon-296 can be explained with the fiducial model 
($\alpha=1$ and $\beta=1$). This value of $\beta$ corresponds to material  
concentrated at the inner edge of the hole, a fact consistent with the small 
period of the variability which locates the extinct material very close to the star.

3.- Either the extinction in the IR caused by the hole material or the stellar
occultations do not notoriously contribute to $\Delta mag_\mathrm{IR}$, ending with a very low value. Note that as $R_\mathrm{H}$ increases, the fraction of material occulted by the star decreases, and therefore its contribution to $\Delta mag_\mathrm{IR}$ also decreases. Thus, Mon-296 is the object that contributes the largest to $\Delta mag_\mathrm{IR}$ as can be seen in the modeling. 
 
4.- The extinction is given by the surface density along the line of sight,
the largest value corresponds to Mon-296, resulting in the largest contribution
to $\Delta mag_\mathrm{IR}$. Thus, along with the smallest $R_\mathrm{H}$ as mentioned in the previous item, both facts leads to the largest non-negligible value of 
$\Delta mag_\mathrm{IR}$ for Mon-296.

5.- According to the modeling, Mon-314, and Mon-433 require an
optically thick inner disk to interpret the lcs, such that we suggest that
instead of hosting a TD they host a PTD. The existence
of this structure helps to increase the magnitude amplitude for
the IR lcs as it is required to interpret the observations. In order to pursue
this idea, in the SED modeling presented by \citet{2019A&A...629..A67} should be
included a small optically thick inner disk. Note that a small size disk is 
enough to produce the occultations required to explain $\Delta mag_\mathrm{obs}$ without
notoriously changing the fitting by \citet{2019A&A...629..A67}.
 
6.- Using the tool Sedfitter based on Hyperion, a new SED fitting is found for Mon-1308. The new fit favors a full-disk instead of a TD. This is consistent with the modeling of the lcs by \citet{2019A&A...625..A45} and our inability to explain the lcs using optically thin material in the hole.
We do not have the tools to test the PTD scenario but also is a possible scenario for Mon 1308. Our final remark is that we cannot be confident with a SED fitting alone, an analysis of lcs in the optical and in the IR can give us relevant information to doubt about this preliminary result, i.e. a revisit of Mon-1308 lead us to conclude that the most probable configuration for Mon-1308 is a full-disk.

\end{document}